\documentclass{elsart}

\usepackage[square,comma]{natbib}
\usepackage{graphicx}
\usepackage{pxfonts}
\usepackage{lineno}

\usepackage{amssymb}

\journal{}

\begin{document}

\thispagestyle{empty}
\begin{Large}
\textbf{DEUTSCHES ELEKTRONEN-SYNCHROTRON}

\textbf{\large{Ein Forschungszentrum der
Helmholtz-Gemeinschaft}\\}
\end{Large}

DESY 10-005

January 2010

\begin{eqnarray}
\nonumber &&\cr \nonumber && \cr \nonumber &&\cr
\end{eqnarray}
\begin{eqnarray}
\nonumber
\end{eqnarray}
\begin{center}
\begin{Large}
\textbf{The potential for extending the spectral range accessible
to the European XFEL down to 0.05 nm}
\end{Large}
\begin{eqnarray}
\nonumber &&\cr \nonumber && \cr
\end{eqnarray}

\begin{large}
Gianluca Geloni,
\end{large}
\textsl{\\European XFEL GmbH, Hamburg}
\begin{large}

Vitali Kocharyan and Evgeni Saldin
\end{large}
\textsl{\\Deutsches Elektronen-Synchrotron DESY, Hamburg}
\begin{eqnarray}
\nonumber
\end{eqnarray}
\begin{eqnarray}
\nonumber
\end{eqnarray}
ISSN 0418-9833
\begin{eqnarray}
\nonumber
\end{eqnarray}
\begin{large}
\textbf{NOTKESTRASSE 85 - 22607 HAMBURG}
\end{large}
\end{center}
\clearpage
\newpage

\begin{frontmatter}



\title{The potential for extending the spectral range accessible to the
European XFEL down to 0.05 nm}


\author[XFEL]{Gianluca Geloni\thanksref{corr},}
\thanks[corr]{Corresponding Author. E-mail address: gianluca.geloni@xfel.eu}
\author[DESY]{Vitali Kocharyan}
\author[DESY]{and Evgeni Saldin}

\address[XFEL]{European XFEL GmbH, Hamburg, Germany}
\address[DESY]{Deutsches Elektronen-Synchrotron (DESY), Hamburg,
Germany}

\begin{abstract}
Specifications of the European XFEL cover a range of wavelengths
down to 0.1 nm. The baseline design of the European XFEL assumes
standard (SASE) FEL mode for production of radiation i.e. only one
photon beam at one fixed wavelength from each baseline undulator
with tunable gap. Recent developments in the field of FEL physics
and technology form a reliable basis for an extensions of the mode
of operation of XFEL facilities. This paper explores how the
wavelength of the output radiation can be decreased well beyond
the European XFEL design, down to $0.05$ nm. In the proposed
scheme, which is based on the use "fresh bunch" technique,
simultaneous operation at two different wavelengths possible. It
is shown that one can generate simultaneously, in the same
baseline undulator with tunable gap, high intensity radiation at
$0.05$ nm at saturation, and high intensity radiation around
$0.15$ nm according to design specifications. We present a
feasibility study and we make exemplifications with the parameters
of SASE2 line of the European XFEL.
\end{abstract}

%
%

\end{frontmatter}



\section{\label{sec:intro} Introduction}

Three X-ray Free-Electron Lasers (XFELs),  LCLS \cite{LCLS},  SCSS
\cite{SCSS}, and the European XFEL \cite{XFEL} are currently under
commissioning or under construction. These machines are based on
the Self-Amplified Spontaneous Emission (SASE) process
\cite{KOND}-\cite{PELL}. Lasing at wavelength as short as $0.15$
nm has been recently demonstrated at LCLS \cite{LCLS2}, together
with operation with electron bunch durations of less than $10$ fs
\cite{DING}. Also, at LCLS saturation has been reached within 20
undulator cells, out of the 33 available. This result has been
enabled by the high-performance beam-formation system at LCLS,
which works as in the ideal operation scenario.

This optimal scenario should be exploited to provide users with
the best possible fruition opportunities. Elsewhere \cite{OUR0} we
suggested to use the extra-undulator length available to provide
two short (sub-ten fs), powerful (ten GW-level) pulses of coherent
x-ray radiation at different wavelengths for pump-probe
experiments at XFELs, with minimal hardware changes to the
baseline setup.

\begin{figure}[tb]
\includegraphics[width=1.0\textwidth]{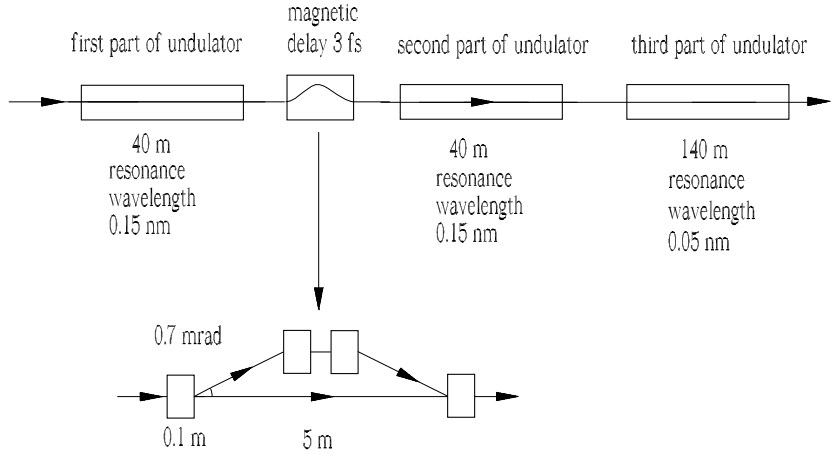}
\caption{Design of the undulator system for the generation of high
power femtosecond SASE pulses at 0.05 nm. The scheme is based on
the use of a fresh bunch technique.  The FEL amplification process
at $0.05$ nm starts up from shot noise.  As a result, the third
part of the undulator must be relatively long, but the hardware is
simpler compared to the self-seeding option discussed in this
paper.} \label{alterna1}
\end{figure}

\begin{figure}[tb]
\includegraphics[width=1.0\textwidth]{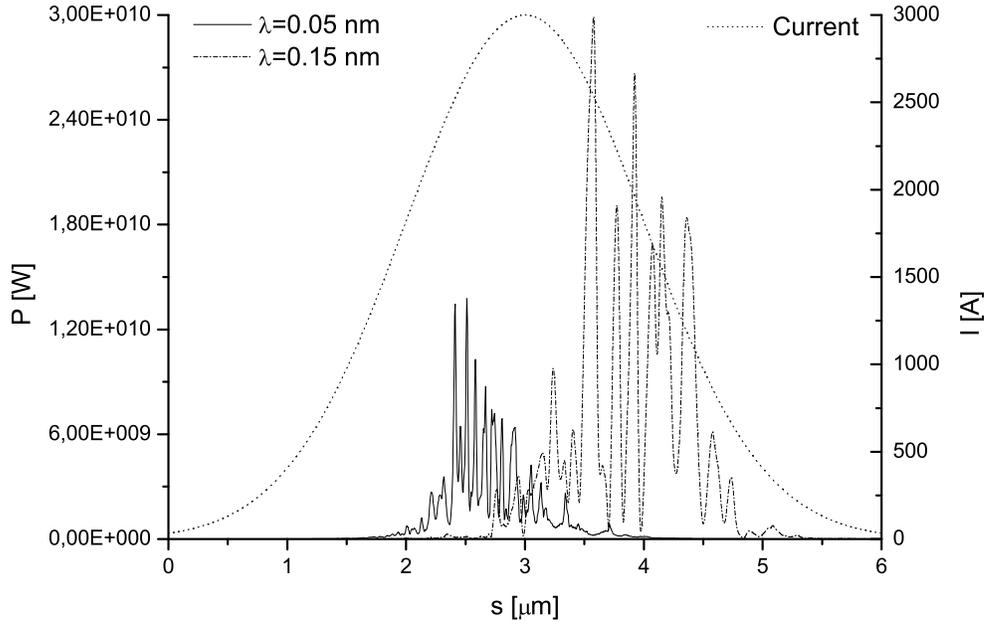}
\caption{Superimposed radiation pulses at $1.5{\AA}$ and $0.5
\AA$. The bunch current is also shown.} \label{alterna2}
\end{figure}
In principle, the scheme proposed in \cite{OUR0} can be easily
extended to generate radiation down to $0.05$ nm. The scheme is
illustrated in Fig. \ref{alterna1}. The first two parts of the
undulator generate radiation at $0.15$ nm, based on a fresh bunch
technique according to the method illustrated in \cite{OUR0}. The
heading half of the bunch is seeded by the radiation produced in
the first part of the undulator, and saturates in the second part,
while the rear part of the bunch enters the linear regime in the
SASE mode. As a result, the rear part of the electron bunch is
still suitable for the SASE process in the third undulator part.
After the second part we still have about $160$ m of undulator
available for SASE2, so that we can reach a high power level at a
different wavelength, in our case $0.05$ nm. In particular, our
calculations are performed for SASE2 assuming that the third part
of the undulator is $26$ modules-long, corresponding to $158.6$ m.
Results are shown in Fig. \ref{alterna2}, which include the
radiation pulse at $0.15$ nm, as discussed in \cite{OUR0} and the
short wavelength emission at $0.05$ nm. Fig. \ref{alterna2}
demonstrates that two femtosecond, ten GW level pulses of coherent
x-rays with two different colors at $0.15$ nm and $0.05$ nm can be
generated with this method.  The advantage of this method is in
the simple hardware required (only a short magnetic chicane) and
in the possibility of independently tuning the wavelengths of both
pulses. However, a fairly long undulator is needed. Note that the
LCLS is equipped with a shorter undulator than SASE2. Therefore,
for LCLS the requirement of a shorter undulator is of crucial
importance, while for SASE2 this option can be practically
realized.

In this paper we propose a scheme for the simultaneous generation
of two maximal-power pulses of radiation around $0.1$ - $0.15$ nm
and at $0.05$ nm, using an undulator which is about $100$ m
shorter than the nominal SASE2 length, so that the extra-available
undulator length can be kept for other purposes. This scheme is
especially important for shorter undulators like for SASE1 and
LCLS, which have fixed gap and can operate at fixed wavelength
only. Technically, the gap of these undulators can be tuned to
$0.05$ nm and fixed once and for all.  For this kind of undulators
our scheme gives the possibility to tune and fix the gap so to
keep the design mode operation at $0.1$ - $0.15$ nm  and add an
extra operation-mode at $0.05$ nm. By this, we answer the interest
of the scientific community towards shorter operation wavelengths,
which would e.g. enable better spatial resolution in elastic x-ray
scattering experiments, providing at the same time less
photoelectric absorption, longer penetration depth, less damage
and larger scattering volume \cite{MASS}.

\begin{figure}
\begin{center}
\includegraphics*[width=100mm]{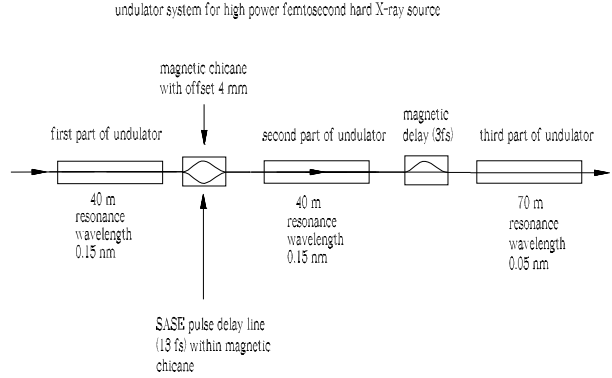}
\caption{\label{sketch2} Undulator system for generation high
power femtosecond SASE pulse at 0.05 nm }
\end{center}
\end{figure}
The overall idea is sketched in Fig. \ref{sketch2}. In the first
part of the undulator, SASE radiation is produced at $0.15$ nm in
the linear regime and in the short-bunch operation mode, according
to the parameters in Table \ref{tab:fel-par}, and described in
more detail in \cite{OUR0}. At the next step, the electron beam
passes through an optical-delay stage composed by a mirror chicane
and a magnetic chicane. The optical delay-stage delays the photon
beam with respect to the electrons of about half of the bunch
length. The magnetic chicane performs two actions. First, it
allows for the installation of the mirror chicane. Second, the
strength of magnetic chicane as dispersion section is sufficient
for suppression of the beam bunching \cite{OUR0}. As a result, the
amplification process in the second undulator section starts with
a "fresh" electron beam, and with radiation produced by the first
undulator section. This is the essence of the "fresh bunch"
technique which was introduced in \cite{HUAYU}-\cite{SAL2}.
Subsequently, in the second part of the undulator the seeded half
of the electron beam saturates within a short distance, while the
non-seeded half of the electron beam lases in the linear regime
only and remains unspoiled. Superimposed to the main
first-harmonic pulse there is a percent-level ($100$ MW-level)
third-harmonic pulse at $0.05$ nm. The third-harmonic component is
superimposed to the non-seeded half of the electron beam by
delaying the electrons with respect to the photon beam with the
help of a magnetic chicane between the second and the third part
of the undulator. Then, the fresh bunch half of the bunch is fed
into the third undulator part together with the first and the
third harmonic pulse, Fig. \ref{sketch2}. The third undulator part
is tuned at $0.05$ nm, so that the first harmonic pulse is not
resonant, while the $100$ MW-level third-harmonic pulse acts as
seed and enables production of a $10$ GW-level pulse at $0.05$ nm
at the exit of the third undulator part.

\begin{table}
\caption{Parameters for the short pulse mode used in this paper.
The undulator parameters are the same of those for the European
XFEL, SASE2, at 17.5 GeV electron energy. }

\begin{small}\begin{tabular}{ l c c}
\hline
& ~ Units &  Short pulse mode \\
\hline
Undulator period      & mm                  & 47.9   \\
Undulator length      & m                   & 256.2  \\
Length of undulator segment        & m                   & 5.0    \\
Length of intersection             & m                   & 1.1    \\
Total number of undulator cells    & -                   & 42     \\
K parameter (rms)     & -                   & 1.201-2.513  \\
$\beta$               & m                   & 17     \\
Wavelength            & nm                  & 0.05 - 0.15   \\
Energy                & GeV                 & 17.5   \\
Charge                & nC                  & 0.025  \\
Bunch length (rms)    & $\mu$m              & 1.0    \\
Normalized emittance  & mm~mrad             & 0.4    \\
Energy spread         & MeV                 & 1.5    \\
\hline
\end{tabular}\end{small}
\label{tab:fel-par}
\end{table}

\section{Feasibility study \label{two}}

In the following we describe the outcomes of computer simulations
using the code Genesis $1.3$ \cite{GENE}.

\subsection{First stage}

\begin{figure}[tb]
\includegraphics[width=1.0\textwidth]{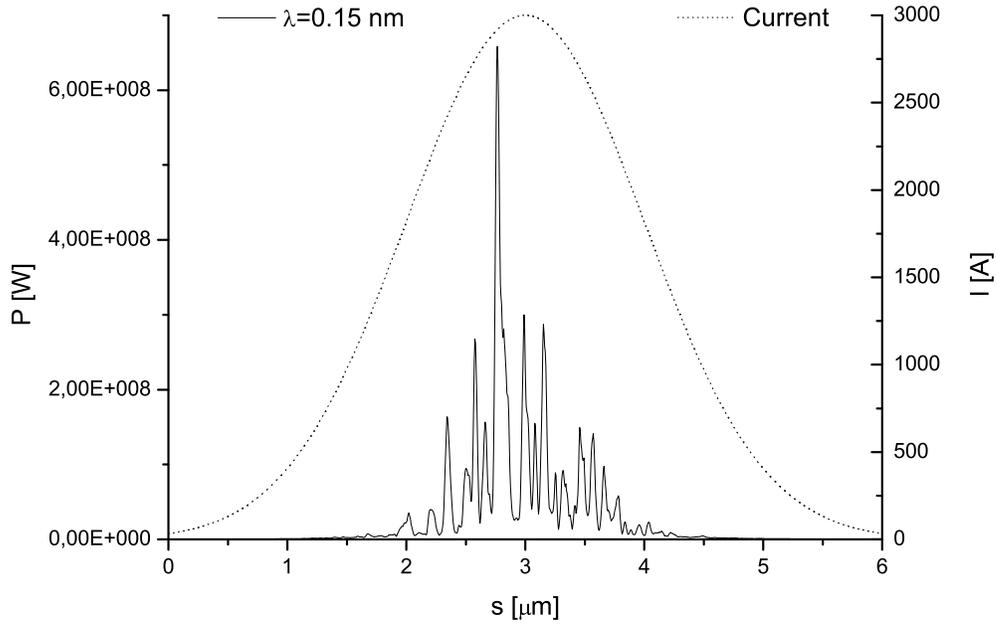}
\caption{Beam power distribution at the end of the first stage
after 7 cells ($42.7$ m). } \label{IstageP}
\end{figure}

\begin{figure}[tb]
\includegraphics[width=0.5\textwidth]{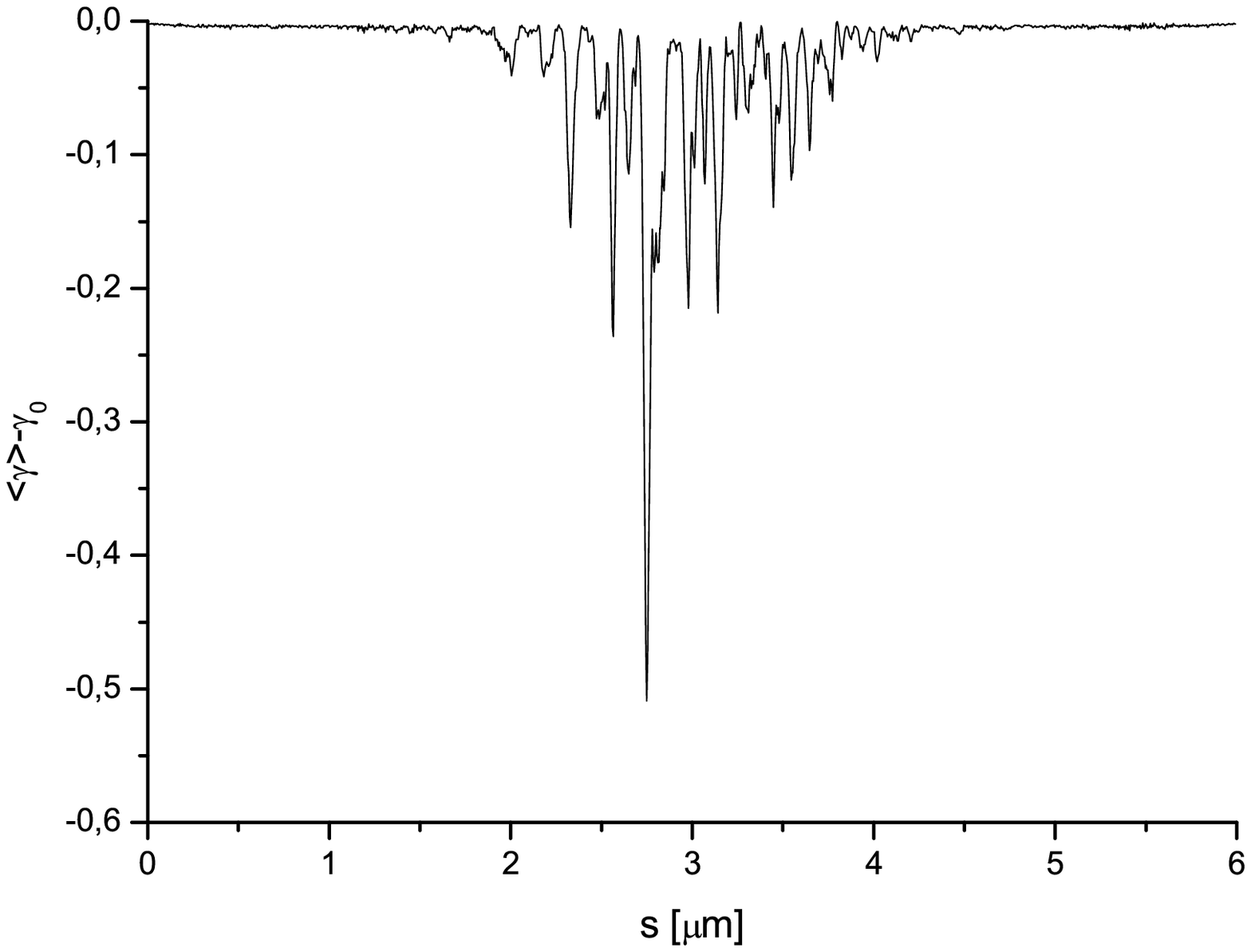}
\includegraphics[width=0.5\textwidth]{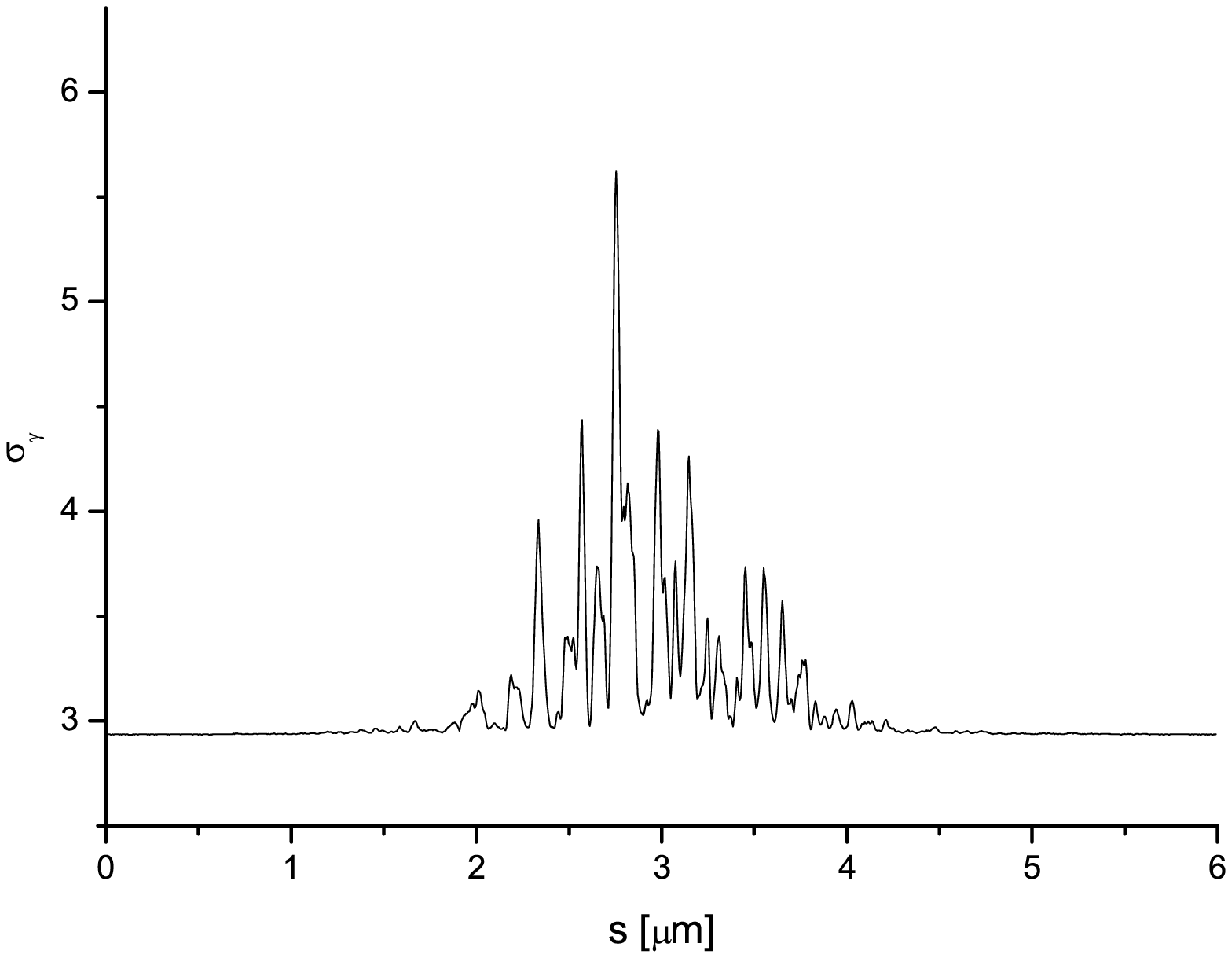}
\caption{Electron beam energy loss (left) and induced energy
spread (right) at the end of the first stage after $7$ cells
($42.7$ m).} \label{Istageen}
\end{figure}

In the first undulator part, which is 7 cells long, each
consisting of a $5$m-long undulator segment and a $1.1$m-long
intersection, for a total length of $42.7$ m. The beam power
distribution is shown in Fig. \ref{IstageP}. The electron beam
energy loss and the induced energy spread are shown in Fig.
\ref{Istageen}. A short (sub $10$ femtosecond) pulse of radiation
is produced in the linear regime ($100$ MW level). Calculations
are identical to the first stage of the pump-probe scheme reported
in \cite{OUR0}.


\subsection{Optical delay}

\begin{figure}
\begin{center}
\includegraphics*[width=100mm]{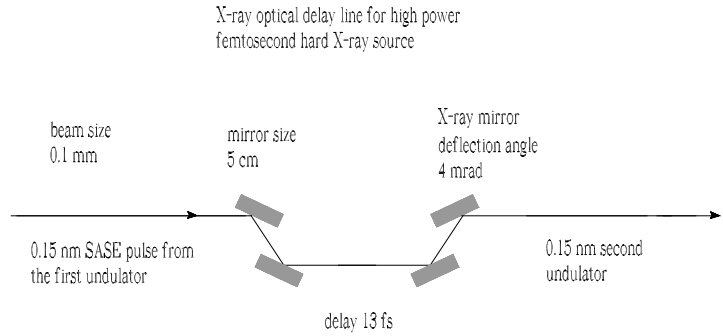}
\caption{\label{sketch3} X-ray optical delay line for high power
hard X-ray source. }
\end{center}
\end{figure}

At variance with respect to the pump-probe scheme in \cite{OUR0},
the present scheme uses from the very beginning an optical delay
stage. This choice is justified by the larger glancing angle for
the radiation at $0.15$ nm, which is about $2$ mrad, with respect
to that at $0.05$ nm, which is $1$ mrad only. The idea is to
install a mirror chicane between the first and the second part of
the undulator, as shown in Fig. \ref{sketch3}. Issues concerning
heat load and size of the mirror have already been discussed in
\cite{OUR0}.

\begin{figure}
\begin{center}
\includegraphics*[width=100mm]{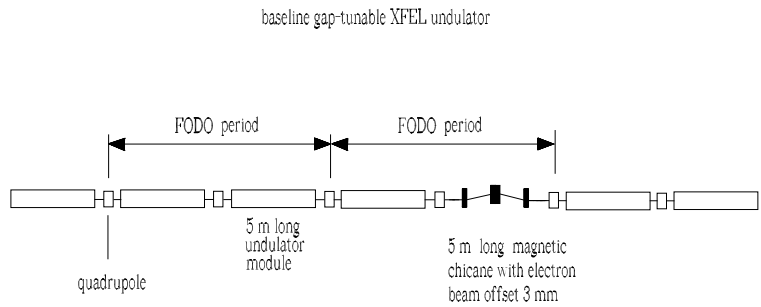}
\caption{\label{sketch0} Installation of a magnetic delay in the
baseline XFEL undulator.  }
\end{center}
\end{figure}

Of course, in order to install the mirror chicane one needs to
first create an offset for the electron trajectory, meaning that a
magnetic chicane should be inserted at the position of the
mirror-chicane, Fig. \ref{sketch0}. Such chicane has also the
function of washing out electron beam density modulations on the
$0.15$ nm-scale. Issues concerning this possibility have also been
discussed in \cite{OUR0}.

\begin{figure}
\begin{center}
\includegraphics*[width=100mm]{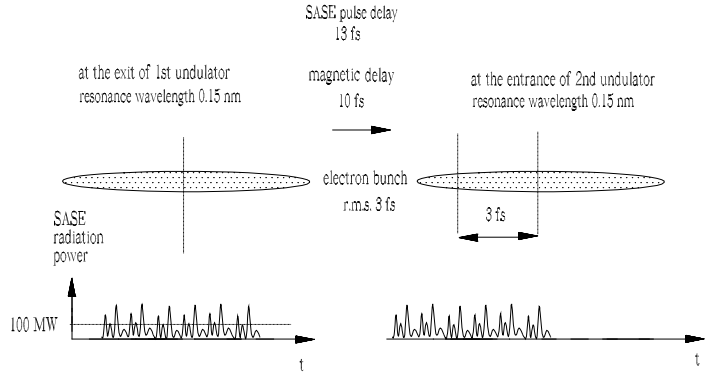}
\caption{\label{sketch4} Sketch of high power 0.05 nm X-ray pulse
generation in the baseline XFEL undulator. First stage: "fresh"
bunch technique. }
\end{center}
\end{figure}

The mirror chicane can be built in such a way to obtain a delay of
the SASE pulse of about $13$ fs. This is enough to compensate a
bunch delay of about $10$ fs from the magnetic chicane, and to
provide the temporal shift in the range $3$ fs, as shown in Fig.
\ref{sketch4}.  It should be noted that the distance between
quadruples in FODO lattice is $6.1$ m, while in our scheme the
length of chicane is $5$ m only, so that the focusing system is
not perturbed.


\begin{figure}[tb]
\includegraphics[width=1.0\textwidth]{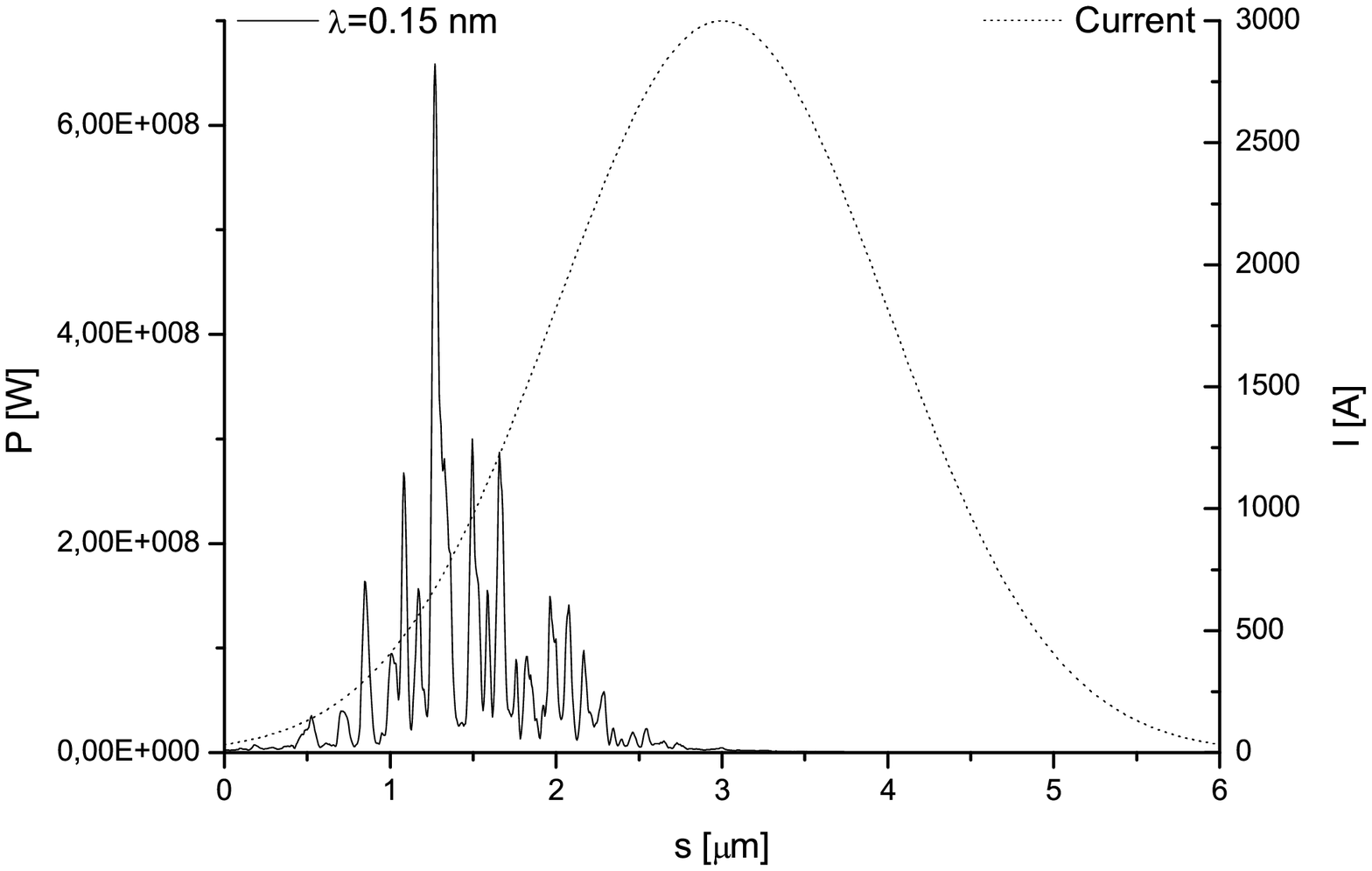}
\caption{Beam power distribution at $0.15$ nm after the optical
delay.} \label{IstagePd}
\end{figure}
As a result of the passage through the magnetic chicane, the
photon beam is delayed with respect to the electron beam of about
$1~\mu$m. The electric field fed into the simulation of the second
part of the undulator is shown in Fig. \ref{IstagePd}. The
electron beam used is generated according to the energy spread and
energy loss distributions in Fig. \ref{Istageen}, similarly as
done in \cite{OUR0}.

\subsection{Second stage}

\begin{figure}[tb]
\includegraphics[width=1.0\textwidth]{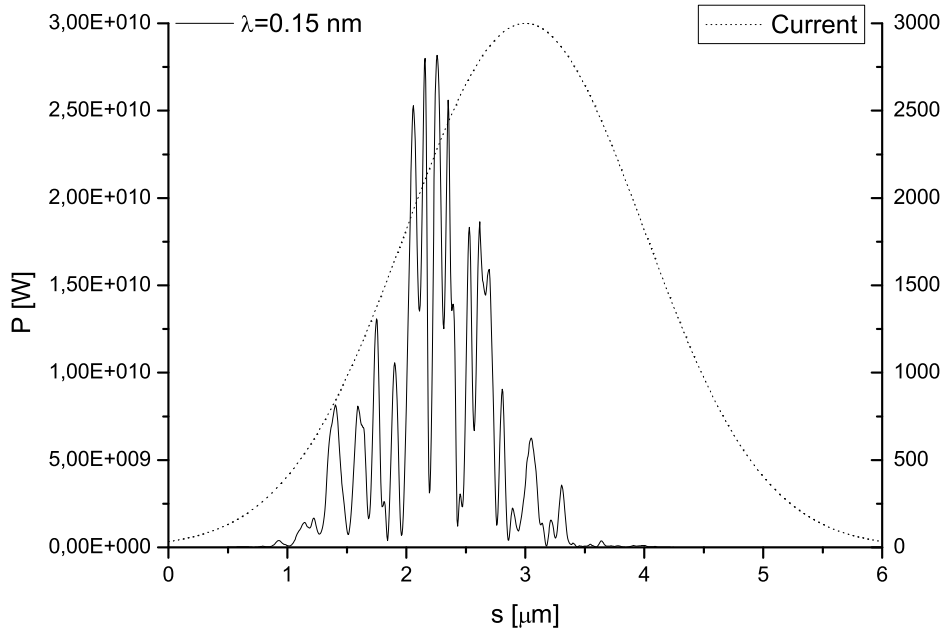}
\caption{Beam power distribution at the end of the second stage,
$7$ cells long ($42.7$ m).} \label{IIstageP}
\end{figure}

\begin{figure}[tb]
\includegraphics[width=1.0\textwidth]{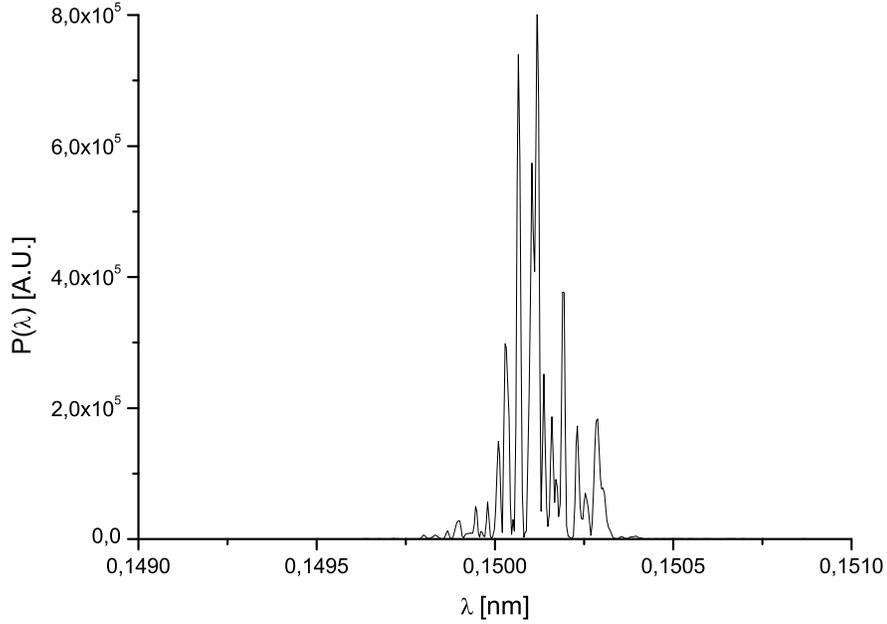}
\caption{Beam power spectrum at the end of the second stage, $7$
cells long ($42.7$ m).} \label{IIstageS}
\end{figure}

\begin{figure}[tb]
\includegraphics[width=0.5\textwidth]{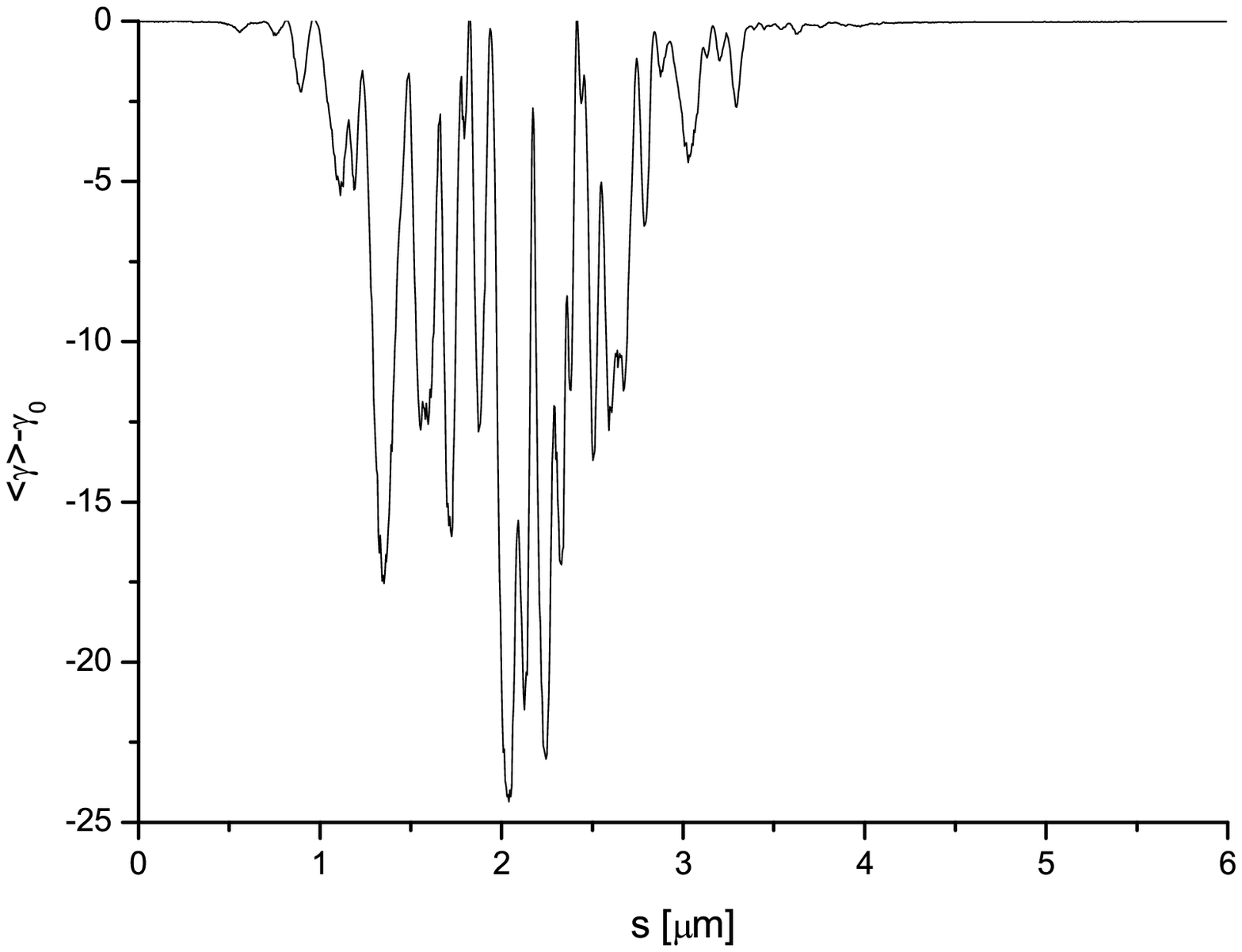}
\includegraphics[width=0.5\textwidth]{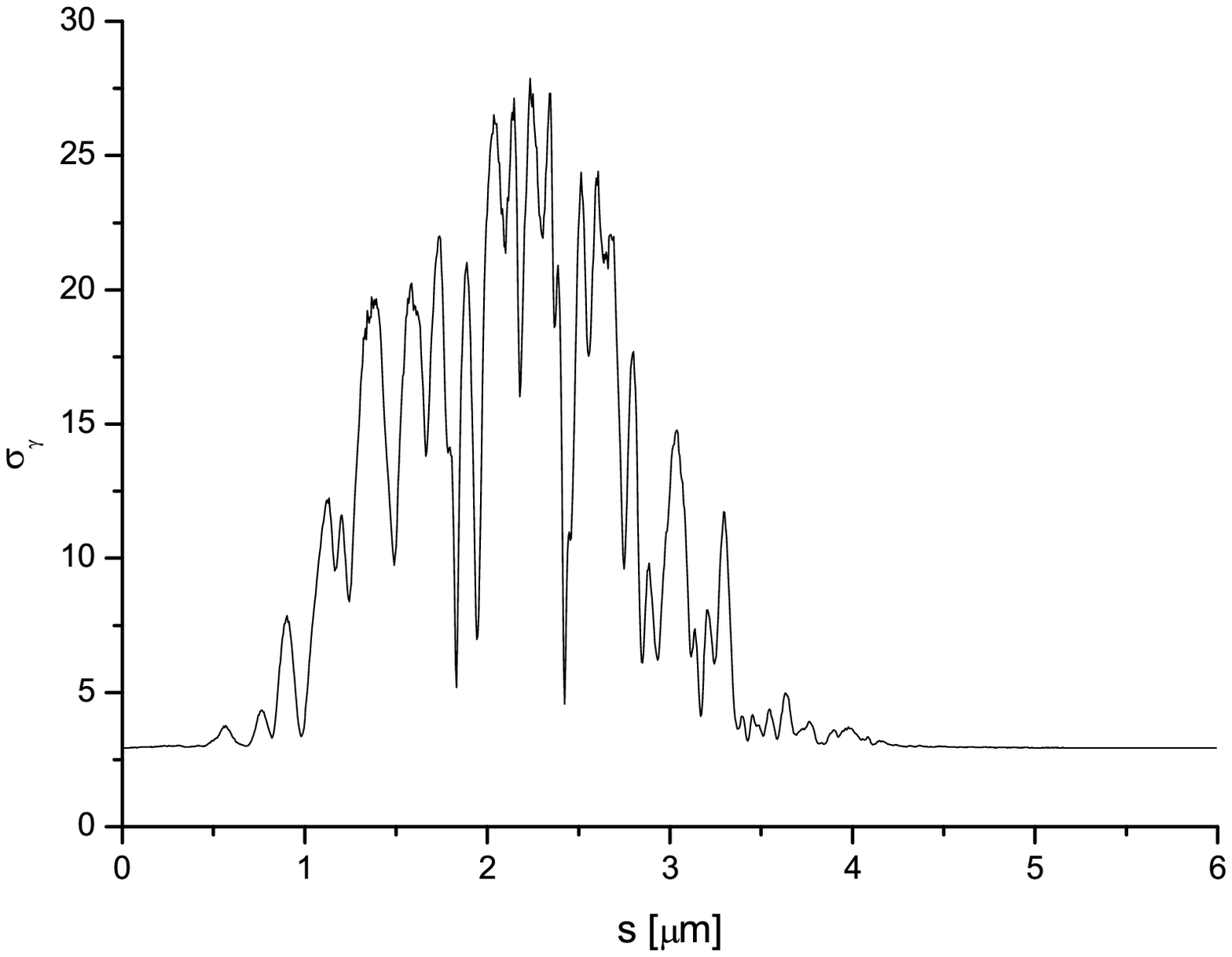}
\caption{Electron beam energy deviation (left) and induced energy
spread (right) at the end of the second stage, $7$ cells long
($42.7$ m).} \label{IIstageen}
\end{figure}

In the second undulator part, the seeded half of the electron
bunch reaches saturation with ten GW power level. The second
undulator part is taken to be another $7$ cells long. The output
power distribution and spectrum is shown in Fig. \ref{IIstageP}
and Fig. \ref{IIstageS}, while energy loss and energy spread are
plotted in Fig. \ref{IIstageen}. The right part of the electron
bunch produces SASE radiation in the linear regime only, which is
negligible.


\subsection{Magnetic delay}

\begin{figure}
\begin{center}
\includegraphics*[width=100mm]{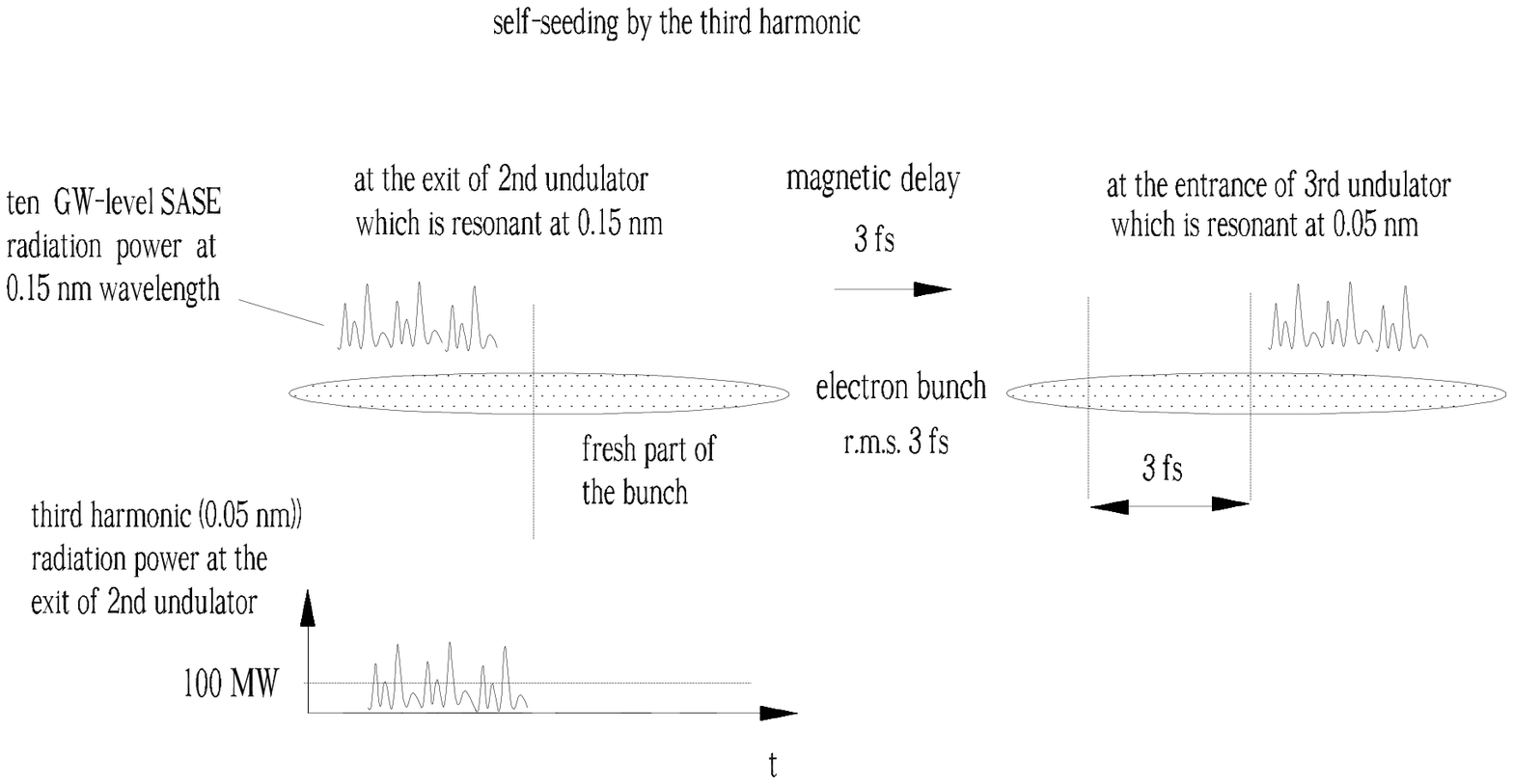}
\caption{\label{sketch} Sketch of third harmonic amplification
from second to third undulator. Ten GW-level on fundamental and
third harmonic can be reached simultaneously  by using combination
fresh bunch technique and self-seeding technique. }
\end{center}
\end{figure}

\begin{figure}[tb]
\includegraphics[width=1.0\textwidth]{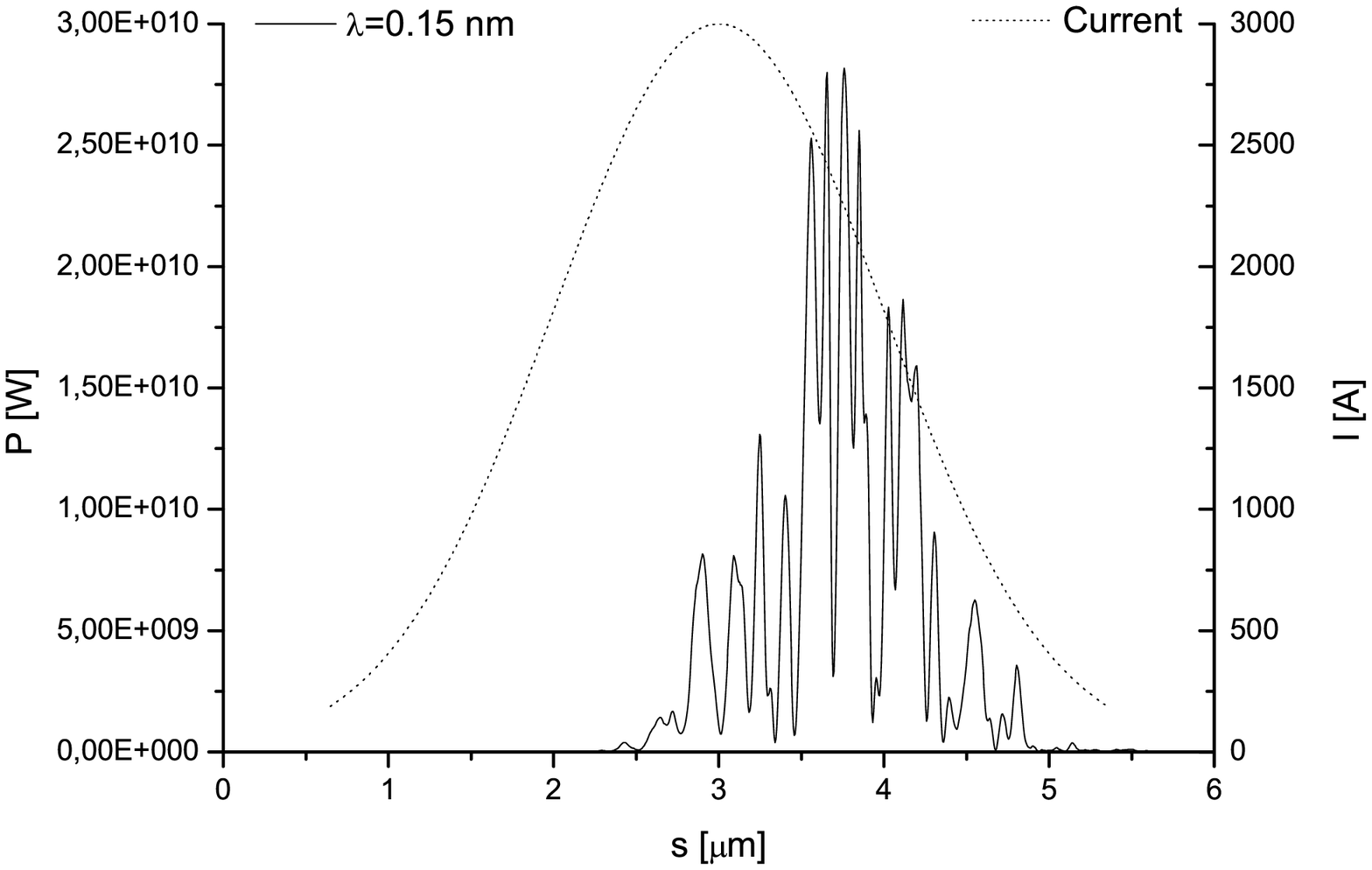}
\caption{Beam power distribution at the entrance of the third
stage, after the magnetic delay, at $0.15$ nm.} \label{firstsh}
\end{figure}

\begin{figure}[tb]
\includegraphics[width=1.0\textwidth]{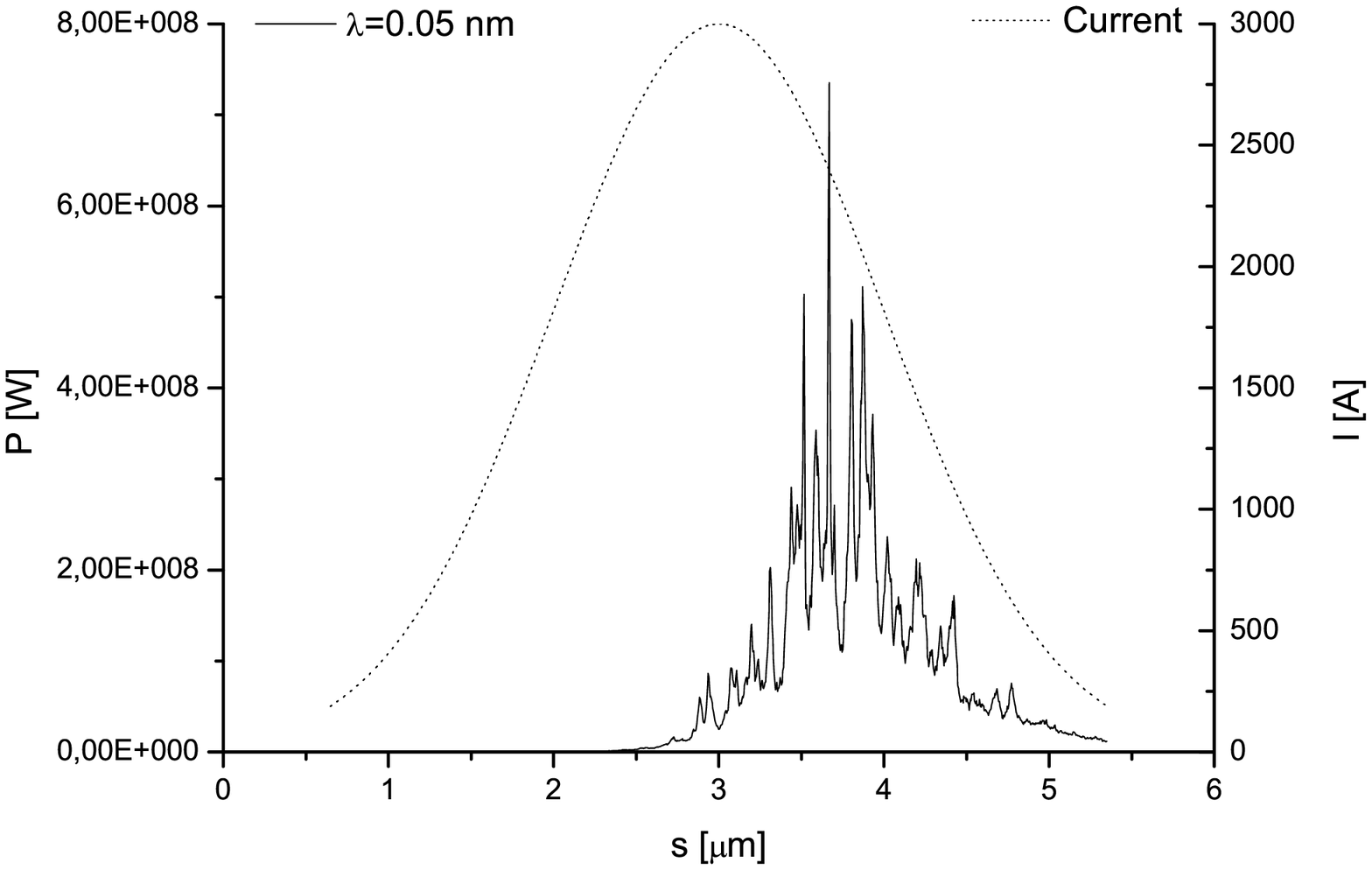}
\caption{Beam power distribution at the entrance of the third
stage, after the magnetic delay, at $0.05$ nm.}
\label{IIIstagePin}
\end{figure}

Together with the saturated first harmonic pulse, the radiation
pulse produced by the seeded (left) half of the electron bunch
also includes some percent-level third harmonic contents. Letting
the bunch through a short chicane, we effectively let the
radiation pulse slip forward\footnote{In these simulations, the
shift amounts to $1.5 \mu$m.}  and seed the fresh (right) half of
the electron bunch, Fig. \ref{sketch}. The effect of the chicane
on the relative position between electrons and photons is shown in
Fig. \ref{firstsh}. First and third harmonic pulses will be
superimposed, but if the third part of the undulator is resonant
with the third harmonic only, the first harmonic will not be of
interest as concerns the SASE process. As a result, in our
simulations, we simulate the magnetic delay by extracting the
third harmonic content and shifted it right with respect to the
electron bunch. The result is shown in Fig. \ref{IIIstagePin}.

\subsection{Third stage}

Following the optical delay, we model the third part of the
undulator, which is $11$ cells-long ($67.1$ m), by feeding the
third harmonic seed in Fig. \ref{IIIstagePin} into the simulation
code. Similarly as before, the electron beam used is generated
according to the energy spread and energy loss distributions in
Fig. \ref{IIstageen}.

\begin{figure}[tb]
\includegraphics[width=1.0\textwidth]{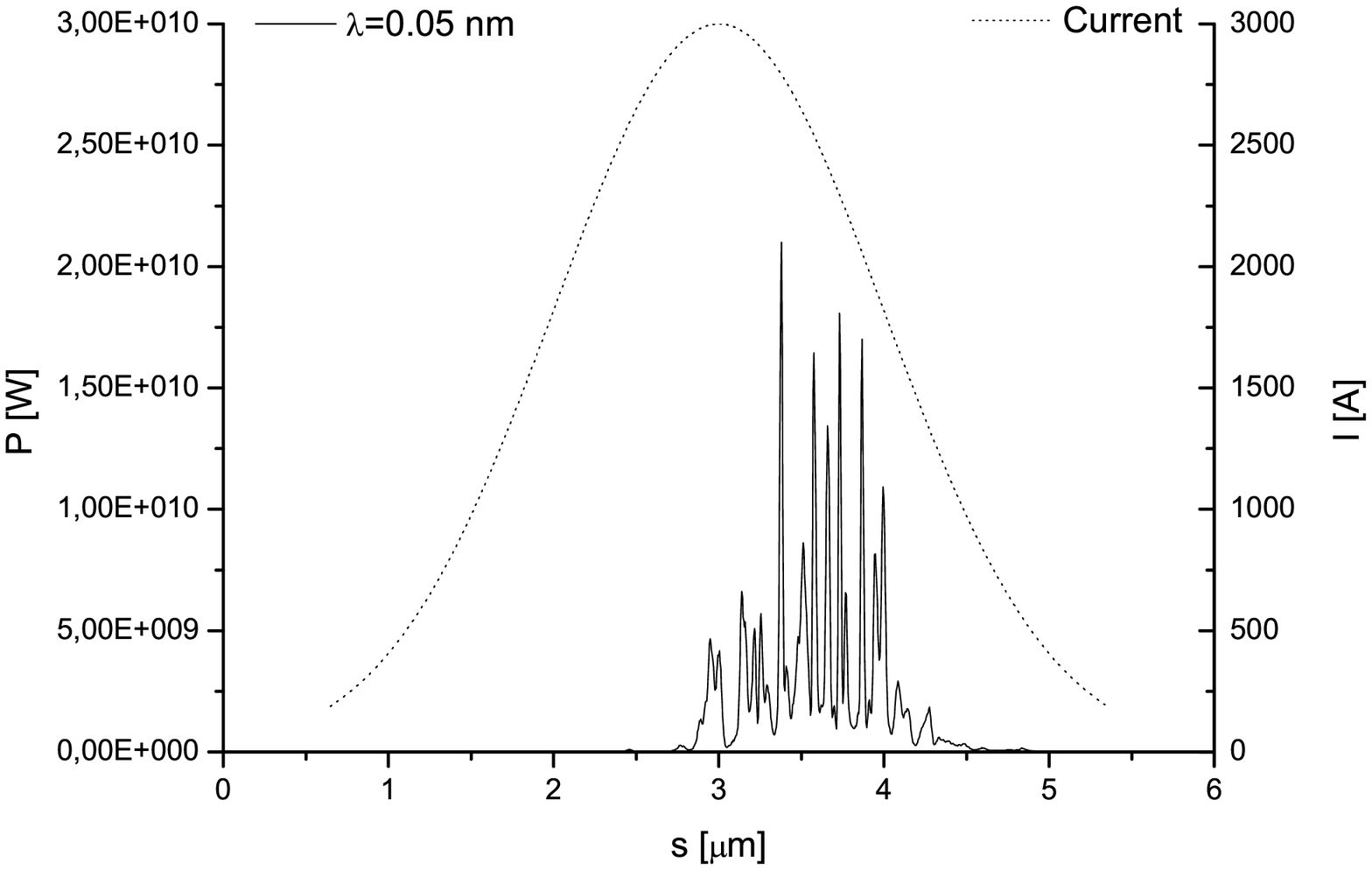}
\caption{Beam power distribution at the end of the third stage,
which is $11$ cells-long ($67.1$ m).} \label{IIIstageP}
\end{figure}
\begin{figure}[tb]
\includegraphics[width=1.0\textwidth]{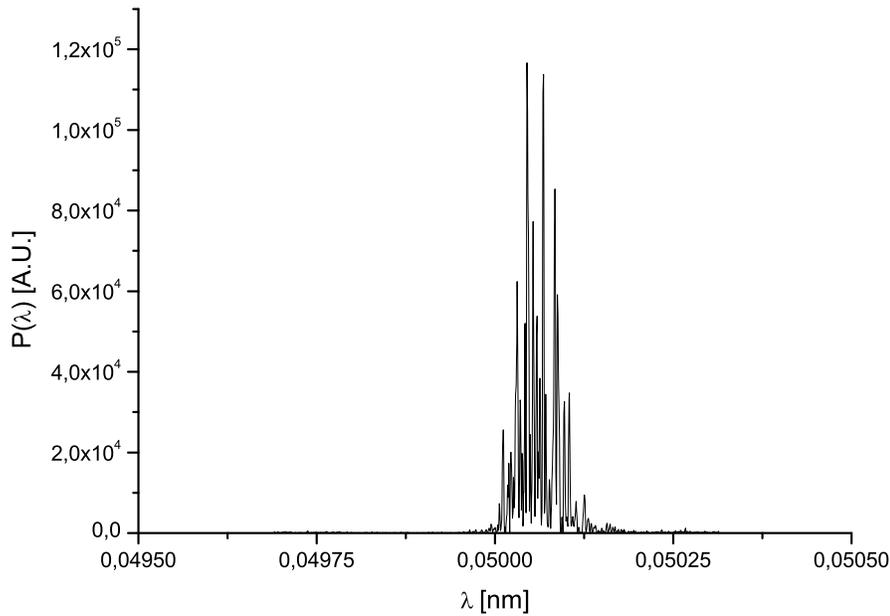}
\caption{Beam power spectrum at the end of the third stage, $11$
cells-long ($67.1$ m).} \label{IIIstageS}
\end{figure}

\begin{figure}[tb]
\includegraphics[width=0.5\textwidth]{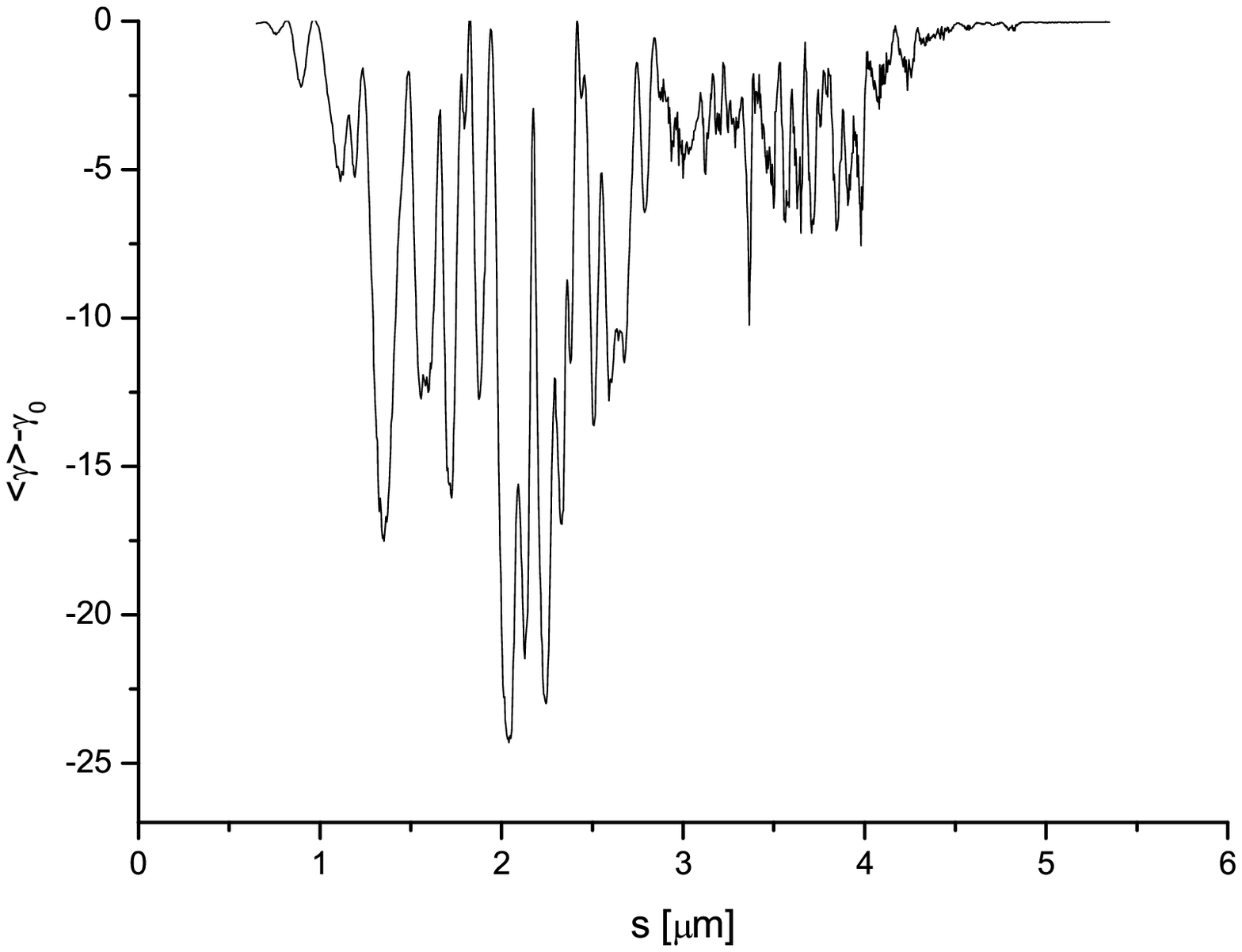}
\includegraphics[width=0.5\textwidth]{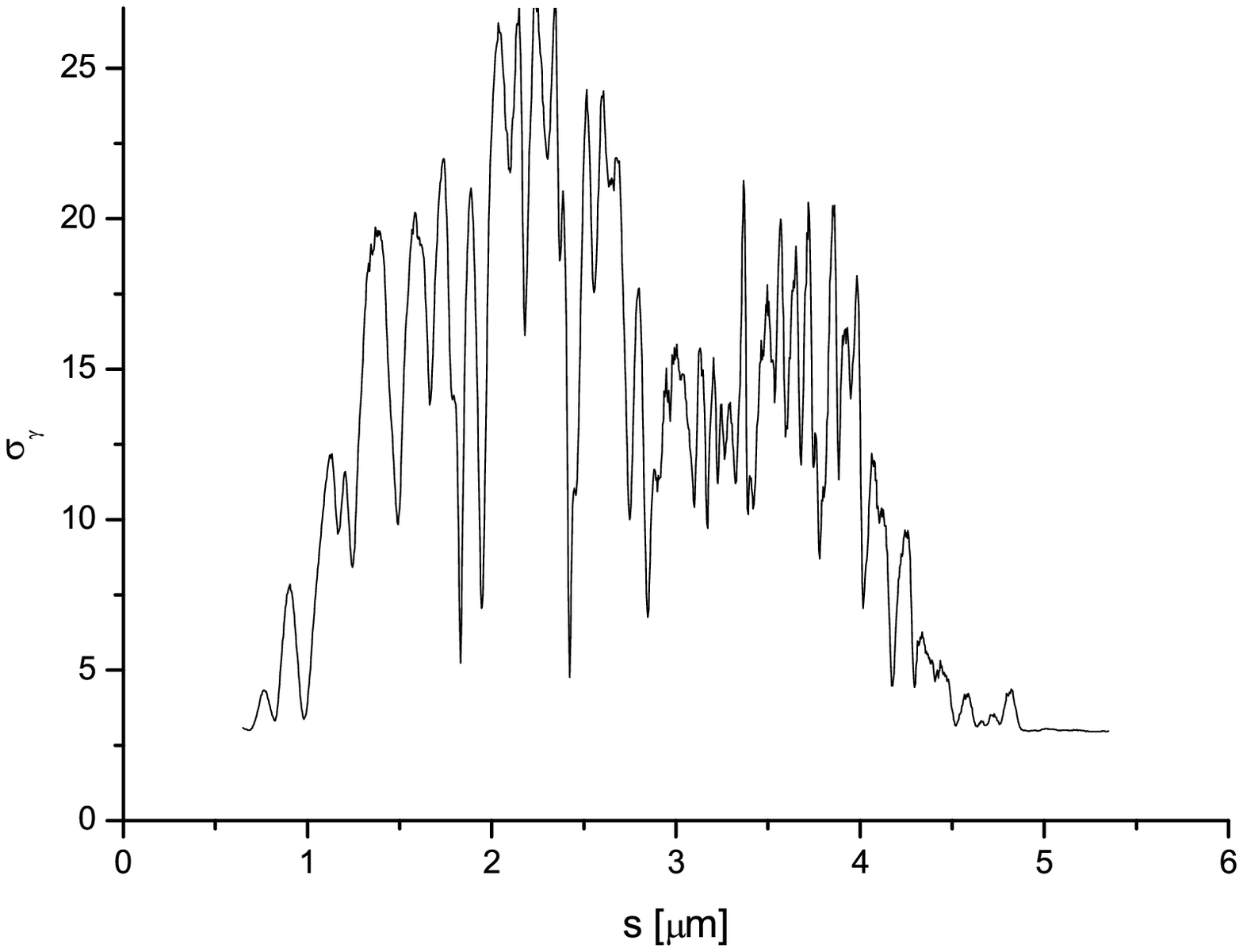}
\caption{Electron beam energy deviation (left) and induced energy
spread (right) at the end of the third stage, which is $11$
cells-long ($67.1$ m).} \label{IIIstageen}
\end{figure}

The output power distribution and spectrum is shown in Fig.
\ref{IIIstageP} and Fig. \ref{IIIstageS}, while energy loss and
energy spread are plotted in Fig. \ref{IIIstageen}.

\subsection{Results}

Fig. \ref{firstsh} and Fig. \ref{IIIstageP} constitute our main
results. Note that the two pulses at $0.15$ nm and $0.05$ nm are
actually superimposed one on top of the other. Like in the case
discussed elsewhere \cite{OUR0}, one faces the task of transport
and utilization of the two radiation pulses to the experimental
station. Transport can be performed with the same optics without
problems, but utilization of the two pulses implies the capability
of separating the two pulses at the experimental station.
Investigating this capability goes beyond the scope of this paper.

\section{Conclusions}

In this paper we propose a technique to extend the baseline
wavelength range of the European XFEL up to $0.05$ nm.
Simultaneously, together with the short-wavelength pulse, a longer
wavelength pulse of similar duration and power is produced at
$0.15$ nm.

Two recent developments in FEL physics have made our technique
feasible. First, the lasing of LCLS and the possibility of working
in the low-charge mode of operation \cite{LCLS2,DING}. Second, the
invention in \cite{OUR0} of a new kind of fresh bunch technique
\cite{HUAYU} allowing for the production of two color pulses with
similar frequency, which relies on a single electron bunch passing
two parts of the same undulator setup, separated with magnetic
delay. Also the scheme proposed in this paper is based on letting
the same electron bunch radiate in separate undulator stages
separated with optical delays and magnetic chicane.

The technique itself is fairly straightforward. Following the
first stage, where seed radiation is produced at $0.15$ nm in the
linear regime, an optical delay stage enables one half of the
electron beam to interact with the seed. After saturation at the
end of the second stage the third harmonic component of the
radiation is further used as a seed in the third stage, which is
resonant at 0.05 nm wavelength. This undulator is long enough (70
m) to reach saturation at the wavelength of 0.05 nm. In the third
undulator the radiation at 0.15 nm plays no role and is diffracted
out of the electron beam.

As the method in \cite{OUR0}, the present method requires very
limited hardware too and is low cost. Moreover, it carries no
risks for the operation of the machine in the baseline mode. Even
though we discuss the case of the European XFEL, our technique may
be taken advantage of by other facilities as well.

\section{Acknowledgements}

We are grateful to Massimo Altarelli, Reinhard Brinkmann, Serguei
Molodtsov and Edgar Weckert for their support and their interest
during the compilation of this work.

\end{document}